\documentclass[
reprint,
superscriptaddress,
amsmath,amssymb,
prx,
]{revtex4-2}

\usepackage{graphicx}
\usepackage{dcolumn}
\usepackage{bm}
\usepackage{hyperref}
\usepackage[dvipsnames]{xcolor}
\usepackage{soul}
\usepackage{ulem}
\usepackage[left]{lineno}
\usepackage{braket}
\begin{document}

\title{Directional transport and nonlinear localization of light in a one-dimensional driven-dissipative photonic lattice}

\author{Tony Mathew Blessan}
\affiliation{Univ. Lille, CNRS, UMR 8523 -- PhLAM -- Physique des Lasers Atomes et Mol\'ecules, F-59000 Lille, France}
\author{Bastián Real}
\affiliation{Departamento de Física, Facultad de Ciencias Físicas y Matemáticas, Universidad de Chile, Santiago, Chile}
\author{Camille Druelle}
\affiliation{Univ. Lille, CNRS, UMR 8523 -- PhLAM -- Physique des Lasers Atomes et Mol\'ecules, F-59000 Lille, France}
\affiliation{Université Jean Monet, F-42023 St-Etienne, France}
\author{Clarisse Fournier}
\affiliation{Univ. Lille, CNRS, UMR 8523 -- PhLAM -- Physique des Lasers Atomes et Mol\'ecules, F-59000 Lille, France}
\author{Alberto Muñoz~de~las~Heras}
\affiliation{Institute of Fundamental Physics, CSIC, Calle Serrano 113b, 28006 Madrid, Spain}
\author{Alejandro~González-Tudela}
\affiliation{Institute of Fundamental Physics, CSIC, Calle Serrano 113b, 28006 Madrid, Spain}
\author{Isabelle~Sagnes}
\affiliation{Université Paris-Saclay, CNRS, Centre de Nanosciences et de Nanotechnologies, 91120 Palaiseau, France}
\author{Abdelmounaim~Harouri}
\affiliation{Université Paris-Saclay, CNRS, Centre de Nanosciences et de Nanotechnologies, 91120 Palaiseau, France}
\author{Luc~Le~Gratiet}
\affiliation{Université Paris-Saclay, CNRS, Centre de Nanosciences et de Nanotechnologies, 91120 Palaiseau, France}
\author{Aristide~Lemaître}
\affiliation{Université Paris-Saclay, CNRS, Centre de Nanosciences et de Nanotechnologies, 91120 Palaiseau, France}
\author{Sylvain~Ravets}
\affiliation{Université Paris-Saclay, CNRS, Centre de Nanosciences et de Nanotechnologies, 91120 Palaiseau, France}
\author{Jacqueline~Bloch}
\affiliation{Université Paris-Saclay, CNRS, Centre de Nanosciences et de Nanotechnologies, 91120 Palaiseau, France}
\author{Clément~Hainaut}
\affiliation{Univ. Lille, CNRS, UMR 8523 -- PhLAM -- Physique des Lasers Atomes et Mol\'ecules, F-59000 Lille, France}
\author{Alberto~Amo}
\email{alberto.amo-garcia@univ-lille.fr}
\affiliation{Univ. Lille, CNRS, UMR 8523 -- PhLAM -- Physique des Lasers Atomes et Mol\'ecules, F-59000 Lille, France}

\begin{abstract}
Photonic lattices facilitate band structure engineering, supporting both localized and extended modes through their geometric design. However, greater control over these modes can be achieved by taking advantage of the interference effect between external drives with precisely tuned phases and photonic modes within the lattice. In this work, we build on this principle to demonstrate optical switching, directed light propagation and site-specific localization in a one-dimensional photonic lattice of coupled microresonators by resonantly driving the system with a coherent field of controlled phase. Importantly, our experimental results provide direct evidence that increased driving power acts as a tuning parameter enabling nonlinear localization at frequencies previously inaccessible in the linear regime. These findings open new avenues for controlling light propagation and localization in lattices with more elaborate band structures. 
\end{abstract}

\maketitle

\section{Introduction}
Photonic lattices serve as an effective platform for studying fundamental wave phenomena while offering precise control over light propagation~\cite{garanovich2012light}. Their geometric arrangement defines the periodic potential experienced by light waves, determining the allowed and forbidden frequency ranges ~\cite{yablonovitch1993photonic, meade2008photonic}, while hopping or coupling interactions between lattice sites enable band structure engineering~\cite{bell2017spectral, Amo2016, ozawa2019topological, Kremer2021}. 

Along with geometric design, interference effects arising from multiple scattering and wave superposition critically shape wave transport in photonic lattices. In the absence of losses, the density of states serves as a key parameter for controlling the transport and localization properties. The design of lattices with bands containing gaps, flat bands, massive or massless dispersions, van Hove singularities or Dirac cones dictates the extent of wave confinement. These spectral modifications give rise to different localization mechanisms including: defect modes, where localized states emerge within the bandgap due to intentional breaking of the translational invariance~\cite{painter1999defect}; Anderson localization, in which random fluctuations suppress wave transport and induce strong spatial confinement~\cite{schwartz2007transport, vatnik2017anderson}; and flat-band states, where destructive interference eliminates dispersion, leading to compact localized states \cite{Vicencio2015, Mukherjee2015a, leykam2018perspective}.

Even though photonic band engineering enables localization in lossless photonic lattices~\cite{Christodoulides2003, garanovich2012light}, its tunability remains inherently limited by the static nature of the system geometry. 
To overcome this constraint, a promising direction is to explore whether localization and wave transport can be controlled through external driving and dissipation.  
In driven-dissipative systems, an external drive, typically a laser, injects light at a given frequency, phase and intensity distribution in the lattice. 
The system evolves toward a steady state dictated by the balance between drive, propagation in the lattice and losses. 
This situation offers a framework in which the response is determined not only by the lattice geometry but also by the driving conditions. 
In particular, when dissipative lattices are driven at resonance, the amplitude and phase of the external driving laser act as tunable parameters that, when tailored to interact with intrinsic lattice modes, can engineer interference effects and, thereby, foster novel transport and localization mechanisms. 
This has been experimentally demonstrated in photonic lattices, where strategic positioning and phase tuning have achieved single-site localization at the micron scale~\cite{jamadi2022reconfigurable}. 
Theoretical studies have also proposed that such control could facilitate unidirectional light transport and sharp localization in extended areas in two-dimensional lattices under laser excitation~\cite{gonzalez-tudela_connecting_2022, de2024nonlinearity, usaj_localization_2024, real2024controlling, salinas_harnessing_2024}.

In the nonlinear regime, the interplay of drive, dissipation and lattice geometry can be tailored to nucleate dissipative solitons~\cite{pernet2022gap, englebert_bloch_2023} and to engineer nonlinear topological modes~\cite{Bardyn2016, ravets_thouless_2025}. 
Engineered dissipation also plays a key role in shaping light-matter interactions at the quantum level. Studies on quantum emitters in structured reservoirs have shown that controlled dissipation enables anisotropic emission and collective effects such as perfect subradiance, where spontaneous emission is strongly suppressed through interference~\cite{pichler_quantum_2015, ramos_non-markovian_2016, gonzalez2017quantum, gonzalez-tudela_anisotropic_2019, gonzalez-tudela_engineering_2019, kannan_-demand_2023, bello_unconventional_2019, leonforte_vacancy-like_2021, kim_quantum_2021, joshi_resonance_2023}.
These findings highlight the impact of interference engineering beyond photonic lattices in the classical regime and extending into quantum optical systems.

\begin{figure*}[t!]
\centering
\includegraphics[width=\textwidth]{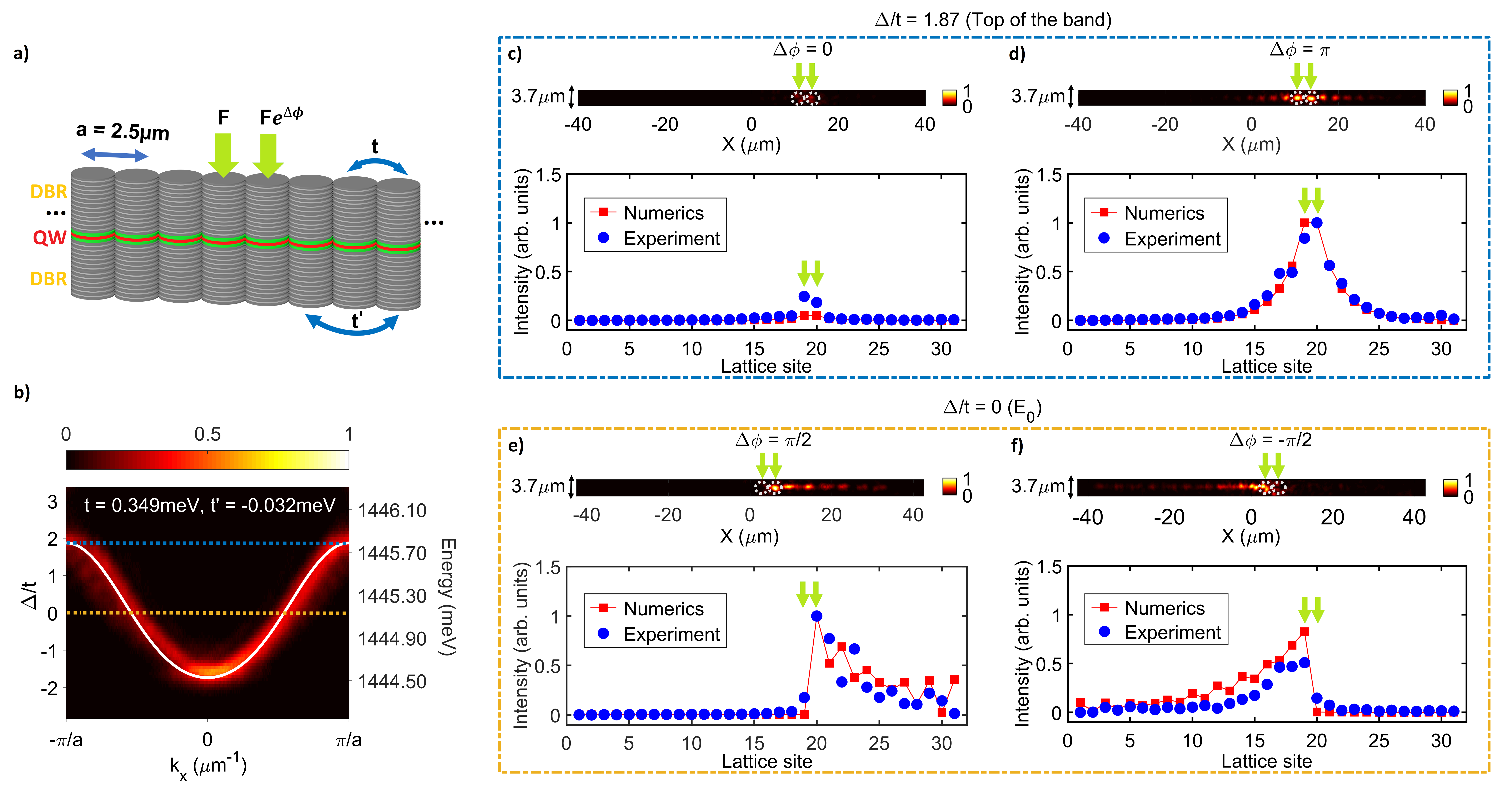} 
\caption{\label{fig:1} (a) Schematic representation of the 1D lattice consisting of 31 micropillars, with a center-to-center separation of \(a = 2.5 \, \mu\text{m}\). The excitation spots are on adjacent pillars, with equal amplitudes \(F\) and phase difference \(\Delta\phi\). 
(b) Angle-resolved photoluminescence measurement of the 1D lattice with one pump spot, showing the energy bands as a function of the in-plane momentum \(k_x\), with \(\Delta = Ep - E_0\), where \(E_0 = 1445.13 \, \text{meV}\). The white line represents the fitted two coupled exciton-photon model. The blue and orange dashed lines indicate the laser energy pumped at the top of the band (\(\Delta/t = 1.87\)) and at zero energy (\(\Delta/t = 0\)), respectively. 
(c, d) Real-space emission and corresponding line profiles for the OFF (\(\Delta\phi = 0\)) and ON state (\(\Delta\phi = \pi\)), both with excitation at the top of the band. 
(e, f) Real-space emission and corresponding line profiles showing rightward propagation (\(\Delta\phi = \pi/2\)) and leftward propagation (\(\Delta\phi = -\pi/2\)), both with excitation at \(\Delta/t = 0\).}
\end{figure*}


In this paper, we present a comprehensive experimental investigation of the directional control of light propagation and the role of nonlinear effects in the localization of light in a one-dimensional lattice of coupled semiconductor micropillars. The simplicity of the one-dimensional lattice compared to previous works in a two-dimensional honeycomb lattice~\cite{jamadi2022reconfigurable} permits a finer control of the experimental parameters to study these effects. 
By precisely tuning the phase of external laser drives and tuning their photon energy to specific band energies, we enable controlled interference effects between the drives and the photonic modes, demonstrating key functionalities such as optical switching, directional propagation, and single-site localization. 
Furthermore, by taking advantage of the Kerr nonlinearity inherent to the semiconductor micropillars in the strong exciton-photon coupling regime, we show that increasing power leads to nonlinearity-driven localization in frequency regimes inaccessible in the linear case as predicted in Refs.~\cite{de2024nonlinearity, usaj_localization_2024}. 
All these phenomena are numerically modeled using a coupled photon-exciton model. Our results provide crucial experimental insights for photonic transport and open new possibilities for tunable light confinement in structured systems. 

\section{Interference between drive and lattice modes}
A one-dimensional lattice of coupled semiconductor micropillars \cite{jacqmin2014direct,baboux2016bosonic,whittaker2018exciton} serves as an ideal system for investigating the interference effects between external drives and lattice modes. 
The micropillars are laterally etched from a $\lambda$ GaAs planar microcavity consisting of two distributed Bragg reflectors of 32 and 36 pairs of $\lambda/4$ layers of Ga$_{0.05}$Al$_{0.95}$As/Ga$_{0.90}$Al$_{0.10}$As, where $\lambda$ is the design wavelength of the cavity at about 858nm. A single InGaAs quantum well is grown at the center of the cavity. At the cryogenic temperature of the experiments ($5$~K), cavity photons and quantum well excitons enter the strong-coupling regime giving rise to exciton polaritons with a measured Rabi splitting of $3.5$~meV. 

Each micropillar undergoes radiative losses to the environment and can be driven by an external laser, forming a driven-dissipative system~\cite{rodriguez2016interaction}. 
The 1D lattice is displayed in Fig.~\ref{fig:1}(a): it has 31 micropillars with a diameter of \(3\,\mu\mathrm{m}\) and center-to-center separations  of \(a=2.5\,\mu\mathrm{m}\). 
The experiments are conducted in transmission geometry with linearly polarized excitation parallel to the lattice direction. 
Photoluminescence or transmission measurement are performed filtering the linear polarization parallel to the lattice and using imaging set-up with a lens of 0.45 numerical aperture and a CCD camera. 
Angle-resolved photoluminescence with a non-resonant laser excitation at 1585.48meV focused on a \(1.1\,\mu\mathrm{m}\) spot (full width at half maximum) centered on top of a micropillar reveals the dispersion relation (Fig.~\ref{fig:1}(b)). The lowest band ($s$-band) follows a cosine-like dispersion, a typical feature of 1D lattices.

The measured band structure is fitted (white line in Fig.~\ref{fig:1}(b)) to a coupled Hamiltonian for the photon-exciton system:
\begin{equation}
H_k = 
\begin{bmatrix}
E_X & \dfrac{\Omega_R}{2} \\ 
\dfrac{\Omega_R}{2} & E_C(k_x)
\end{bmatrix},
\label{eq:hamiltonian}
\end{equation}
where \(E_X\) represents the exciton energy, \(\Omega_R\) is the Rabi splitting indicating the coupling strength between the photon and exciton modes, and \(E_C(k_x)\) is the cavity photon energy. 
We assume the exciton energy to be independent of $k$ due to its large mass of $\sim0.5$~m$_0$ compared to the photon mass in the cavity of $\sim10^{-5}$~m$_0$, with m$_0$ the free electron mass.
Employing a tight-binding description in which photons hop between adjacent micropillars, the photon dispersion of the lowest energy lattice band is given by:
\begin{align}
E_C(k_x) = E_{C}^{0} - 2t\cos(k_x) - 2t'\cos(2k_x)  
\label{eq:disp}
\end{align}
where \(t\) denotes the nearest-neighbor hopping, \(t'\) the next-nearest-neighbor hopping and $E_{C}^{0}$ the bare photon energy of an individual micropillar. 
The next-nearest-neighbor hopping appears in our system as an effective interaction mediated by the hybridization between $s$ and $p$ modes. This type of coupling mechanism has been explored in photonic lattices and polaritonic systems~\cite{mangussi2020multi}.
Using the \texttt{fmincon} function in MATLAB by setting \(E_X\ = 1450.49 \,\mathrm{meV}\) and \(\Omega_R\ = 3.5\,\mathrm{meV}\), which can be deduced from the photoluminescence spectra of the lattice, the optimization results yield \(E_{C}^{0} = 1445.68\,\mathrm{meV}\), \(t = 0.35\pm 0.01\,\mathrm{meV}\) and \(t' = -0.03\pm 0.01\,\mathrm{meV}\).  
We operate at a photon-exciton detuning of -5.45~meV, ensuring that the polaritons are  92\% photonic and 8\% excitonic. 
The laser detuning is defined by \( \Delta = E_p - E_0 \) such that $\Delta=0$ is reached when the pump energy $E_p$ is resonant with \( E_0 = 1445.13\,\mathrm{meV}\), an energy reference close to the middle of the band. 

As a first step to study the interference between the eigenmodes of the lattice and multiple external drives, we pump two adjacent sites (\( m \) and \( m+1 \)), as shown in Fig.~\ref{fig:1}(a), with equal amplitude at a photon energy corresponding to \( \Delta/t = 1.87 \) (top of the band). 
The phase difference \( \Delta \phi \) between the two spots is experimentally set using a piezo-controlled mirror, which adjusts the path length of one beam relative to the other. 
When the two sites are pumped with zero phase difference (\( \Delta \phi = 0 \)), the system exhibits a non-emissive (“OFF”) response, as shown in Fig. 1(c): only a small amount of light is observed at the pumped sites, while the rest of the lattice remains dark. 
The dark response of this driving profile can be understood by considering the spatial antisymmetric shape of the anti-bonding lattice eigenmode at the top of the band. The two laser spots excite coherently and in phase the top band mode at adjacent micropillars. 
Due to the phase difference of \( \pi \) between sites of the top band mode, the injected fields interfere destructively, resulting in a dark response. 
The observed residual transmission at the pumped sites (green arrows in Fig.~\ref{fig:1}(c)) arises from slight misalignment of the excitation beams.

Efficient excitation of the lattice mode at the top of the band requires a relative phase of \( \pi \) between neighboring sites. 
In this case, shown in Fig.~\ref{fig:1}(d), the field injected by the two laser spots interferes constructively and gives rise to a highly emissive (“ON”) state that propagates away from the pump spots. 
If instead of the modes on the top of the band, of antisymmetric nature, we had addressed the symmetric modes of the bottom of the band, we would expect the drive phase pattern of the “ON” and “OFF” response to be reversed.

The observed decay of the emitted intensity away from the pumped sites arises from the continuous escape of photons as they propagate in the lattice.
Note that in these experiments under resonant injection, energy relaxation of the injected photons is negligible.
The measured output field is emitted at the exact same frequency of the driving laser.

\section{Directional Propagation}
Another striking phenomenon that arises from the interference between the drive and lattice modes is directional propagation of light~\cite{real2024controlling}. 
This phenomenon has been thoroughly studied in the context of quantum emitters coupled to waveguides and lattices~\cite{pichler_quantum_2015, ramos_non-markovian_2016, kannan_-demand_2023, joshi_resonance_2023,  gonzalez-tudela_engineering_2019}.
This situation can be observed when setting the drive photon energy to \(\Delta/t = 0 \) (\(E_p=E_0\)) and properly adjusting the phase between two adjacent pump spots. 
Figure~\ref{fig:1}(e, f) displays this directional transport phenomenon for phase differences of \(-\frac{\pi}{2}\) and \(\frac{\pi}{2}\) between the excitation spots. 
When \( \Delta \phi = \frac{\pi}{2} \) (panel (e)), the phase gradient introduced by the pump matches the symmetry of modes with positive group velocity at \( \Delta/t = 0 \), which propagate to the right.
Conversely, for \( \Delta\phi = -\frac{\pi}{2} \) (panel (f)), the reversed phase gradient favors coupling to modes with group velocity directed to the left.
The directionality can be quantitatively evaluated (see Supplemental material Ref.~\cite{Supplementary_arxiv} for further details): for \( \Delta\phi = -\frac{\pi}{2} \), \( 78\% \) of the injected light is directed to the left of the pumped region (excluding the pumped sites), whereas only \( 22\% \) is observed on the right. In contrast, for \( \Delta\phi = +\frac{\pi}{2} \), the directionality is reversed, with \( 92\% \) of the light appearing to the right of the pump spots and just \(8\% \) to the left.

In Fig.~\ref{fig:1}(e), the decaying intensity to the right of the pump spots displays an oscillating behavior. The reason is that the pump spots (\(m = 19, 20\)) are positioned relatively near the edge of the lattice (\(m = 31\)), and the field propagating in the lattice is reflected at the boundary resulting in an interference. In Fig.~\ref{fig:1}(f) this effect is less prominent because the edge \(m = 1\) is located farther from the pumping sites.
Similar results are found with any other choice of pair of pumped sites as long as they are sufficiently far from the edges.

\section{Numerical and analytical models}
To better understand the system and enable a direct comparison with the experiment, we have developed a numerical model of an array of 31 coupled sites in the tight-binding limit. To keep track of the hybrid light matter nature of the micropillar resonances, we model each site \( m \) with a discrete cavity mode \( \psi_{C,m}(t) \) and an exciton mode \( \psi_{X,m}(t) \). Keeping track of the two coupled fields provides deeper insights into the saturation of polariton nonlinearities, and goes beyond previous works on the study of nonlinear localization effects~\cite{de2024nonlinearity, usaj_localization_2024}. These fields are assembled into a state vector \( \mathbf{y}(t) \in \mathbb{C}^{62} \):
\[
\mathbf{y}(t) = 
\begin{bmatrix}
\psi_{C,1}(t) \\
\psi_{X,1}(t) \\
\psi_{C,2}(t) \\
\psi_{X,2}(t) \\
\vdots \\
\psi_{C,31}(t) \\
\psi_{X,31}(t)
\end{bmatrix}.
\]
The time evolution of the system is governed by the following coupled differential equations~\cite{Carusotto2004}:
\begin{align}
i\hbar \frac{d\psi_{C,m}}{dt} &= (\delta_C - i \frac{\gamma_C}{2})\psi_{C,m} 
+ \frac{\Omega_R}{2} \psi_{X,m} \label{eq:photon} \\[1.5ex] \nonumber 
&\quad + t \sum_{\langle n \rangle} \psi_{C,n}
+ t' \sum_{\langle\langle n \rangle\rangle} \psi_{C,n}
+ F_m
\end{align}

\begin{align}    
i\hbar \frac{d\psi_{X,m}}{dt} &= (\delta_X - i \frac{\gamma_X}{2}) \psi_{X,m}
+ \frac{\Omega_R}{2} \psi_{C,m} \label{eq:exciton} \\[1.5ex] \nonumber 
&\quad \quad \quad \quad \quad \quad \quad \quad  +g_X |\psi_{X,m}|^2 \psi_{X,m}.
\end{align}
$\delta_C=E_p-E_{C}^{0}$ and $\delta_X=E_p-E_X$ denote the photon and exciton detunings relative to the pump energy \( E_p \),  the photon and exciton decay rates are \( \gamma_C \) and \( \gamma_X \), the term \( g_X |\psi_{X,m}|^2\) captures the exciton-exciton nonlinearity, and $\langle n \rangle$ and $ \langle\langle n \rangle\rangle$ refer to the nearest- and next-nearest-neighbors. The pump \( F_m \) is applied at two sites \( m_1 \) and \( m_2 \) with a relative phase difference \( \Delta \phi \) as: 
\[
F_m = F_p 
\left( \delta_{m,m_1} + e^{i \Delta \phi} \delta_{m,m_2} \right)
\]
where \(\delta_{m,m_{1,2}}\) is the Kronecker delta function. In the model, only the photon field \( \psi_{C,m} \) undergoes hopping between sites via the nearest (\( t \)) and next-nearest-neighbor (\( t' \)) couplings, while the exciton field \( \psi_{X,m} \) remains localized due to its significantly larger effective mass compared to the photon. At low input intensities, Eqs.~(\ref{eq:photon}) and (\ref{eq:exciton}) have a steady-state solution for each configuration of the drive field \( F_m \). 

We have performed numerical simulations under the conditions of the experiments shown in Fig.~\ref{fig:1}(c)-(f). 
In these simulations in the linear regime, $g_X = 0$ and both excitation spots are driven with equal amplitudes \(F_p\). 
For convenience, we set $\gamma_x=\gamma_c \equiv\gamma$ with a value of \(0.12\,\mathrm{meV}\), which fits the measured decay of the polariton intensity in Fig.~\ref{fig:1}(d). 
We numerically reproduce the key experimental observations: the existence of ON and OFF states in Fig.~\ref{fig:1}(c) and~\ref{fig:1}(d), as well as the directional propagation across the lattice displayed in Fig.~\ref{fig:1}(e) and ~\ref{fig:1}(f)). 
In Fig.~\ref{fig:1}(c), the driven sites exhibit significantly less intensity in the numerical simulations compared to the experiment, which can be attributed to the ideal mode matching between the pump and each site in the simulations. 
The simulation in Fig.~\ref{fig:1}(e) reproduces the intensity oscillations observed in the experiment, which arise from the interference between the right propagating field and the polaritons reflected from the edge of the lattice.

After experimentally observing and numerically confirming scenarios enabling light transport and manipulation in the lattice, we next ask whether additional, less intuitive transport regimes may exist within the accessible parameter space. 
To explore this, we derive an analytical expression based on a Fourier-space solution of the steady-state coupled Eqs.~(\ref{eq:photon}) and (\ref{eq:exciton}) following the methods introduced in Ref.~\cite{de2024nonlinearity}. Specifically, we consider an infinite 1D lattice for two coherent pumps applied at distinct lattice sites \( m_1 \) and \( m_2 \) and solve the equations in momentum space, followed by an inverse Fourier transform to recover the photonic spatial field distribution \( |\psi_{C,m}|^2 \). To simplify the analytical expression, we consider the case \(t'=0\) and the limit of purely photonic polaritons, which implies that the lattice bandwidth is $4t$. The detailed derivation is given in the Supplemental Material~\cite{Supplementary_arxiv}. The photon intensity at each site $m$ takes the form:
\begin{widetext}
\begin{equation}
\displaystyle |\psi_{C,m}|^2 = |F_p|^2 D(\Delta)^2 
\left| 
e^{\displaystyle i k_0 |m - m_1|} e^{\displaystyle - \gamma D(\Delta) |m - m_1|} 
+ e^{\displaystyle i \Delta \phi} e^{\displaystyle i k_0 |m - m_2|} e^{\displaystyle - \gamma D(\Delta) |m - m_2|} 
\right|^2,
\label{eq:analytical}
\end{equation}
\end{widetext}
where $k_0 \equiv \arccos\left(\frac{-\Delta}{2t} \right)$ and $D(\Delta) = \frac{1}{\sqrt{4t^2 - \Delta^2}}$ is the density of states.
From Eq.~(\ref{eq:analytical}), we derive the general conditions that establish the relation between \( \Delta \) and \( \Delta\phi \) for directional propagation to occur. For propagation to the right, we impose destructive interference in the region to the left of the pump spots, i.e. $|\psi_{C,m}|^2=0$ for $ m < \min(m_1, m_2)$ in the limit of negligible losses $\gamma \ll t$. This leads to the condition~\cite{Supplementary_arxiv}:
\[
\cos\left( \Delta \phi + k_0 (m_2 - m_1) \right) = -1
\]
which yields the phase-matching requirement:
\begin{equation}
\Delta \phi + k_0 (m_2 - m_1) = (2\ell + 1)\pi, \quad \ell \in \mathbb{Z}.
\label{eq:right}
\end{equation}
For directional propagation to the left, we get the condition:
\begin{equation}
\Delta \phi - k_0 (m_2 - m_1) = (2\ell + 1)\pi, \quad \ell \in \mathbb{Z}.
\label{eq:left}
\end{equation}

Equations (\ref{eq:right}) and (\ref{eq:left}) reveal that directional transport is tunable: for any value of detuning \( \Delta \) within the band, there exists a corresponding phase difference \( \Delta \phi \) that enables directional propagation, except at the band edges. The appropriate phase difference can be found for any separation between the two pumping spots. In the specific case of \( \Delta = 0 \) (i.e., \( k_0 = \pm \pi/2 \)) and \(m_2 - m_1 = 1\), directional propagation occurs at \( \Delta \phi = \pm \pi/2 \). Our experiments and numerical simulations confirm this situation, even though they deviate slightly from the conditions in which Eqs.~(\ref{eq:analytical})-(\ref{eq:left}) have been obtained: experiments and simulations are done in presence of losses and include weak next-nearest neighbor coupling and a small excitonic component which modify the shape of the pure photonic band with just next-neighbor hoppings assumed in Eq.~(\ref{eq:analytical}).

\section{Localization in linear regime}
Engineering of drive patterns in a lattice can also lead to steady-states localized at a single and several sites~\cite{gonzalez2017quantum, jamadi2022reconfigurable, de2024nonlinearity, usaj_localization_2024}. 
This situation has been investigated in a honeycomb lattice of coupled micropillars, in which a laser drive configuration arranged in three sites leads to the confinement of light onto a single site~\cite{jamadi2022reconfigurable}. 
In this section, we show localization in a 1D lattice as a first step towards investigating nonlinear effects.

\begin{figure}[t!]
\centering
\includegraphics[width=\columnwidth]{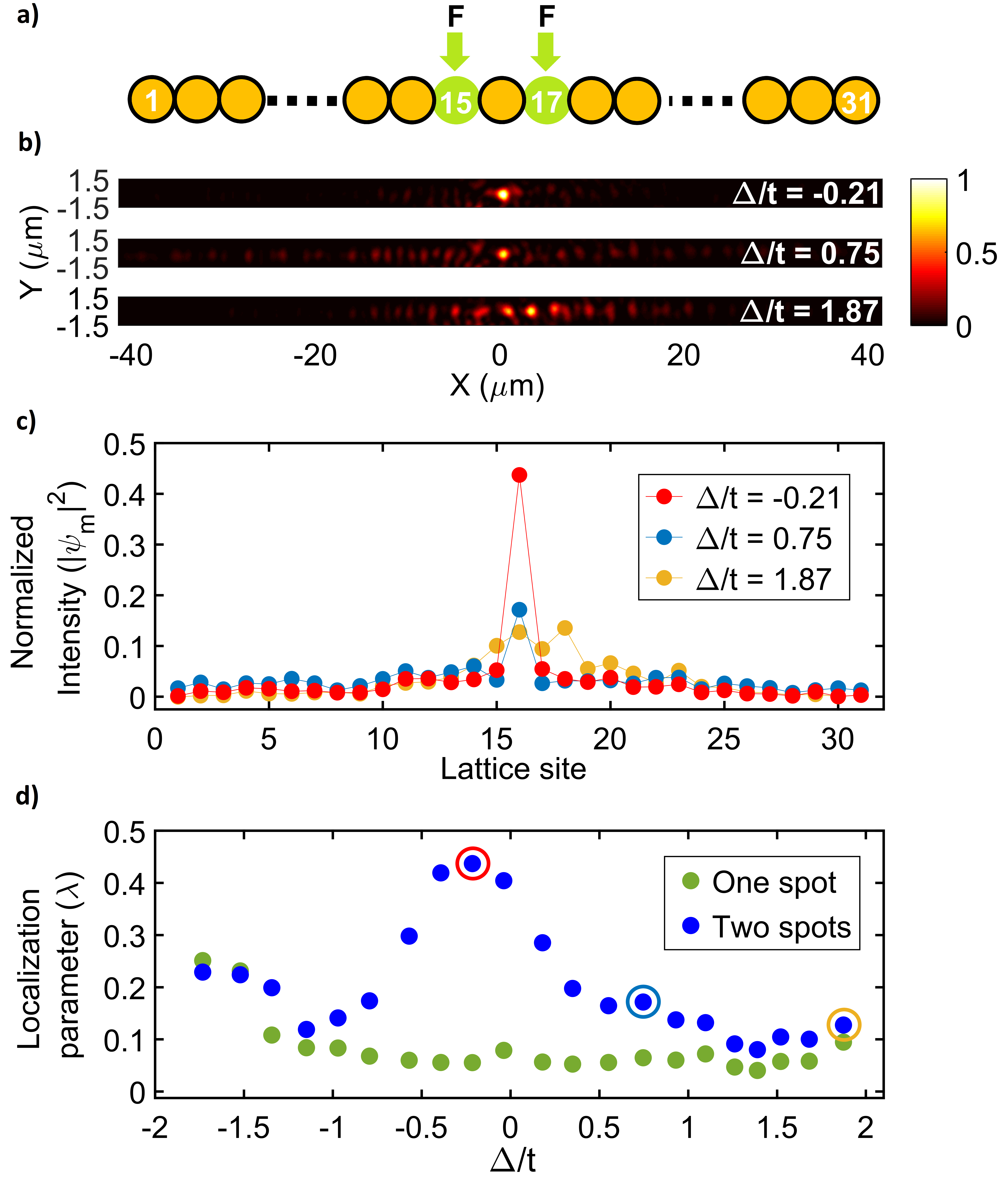}
\caption{(a) Schematic of the 1D lattice consisting of 31 micropillars, with excitation spots with equal phase \(\Delta\phi = 0\) and enveloping a single site. 
(b, c) Real-space emission for three values of \(\Delta/t = -0.21\), \(0.75\), and \(1.87\), along with their corresponding line profiles. 
(d) \(\Delta/t\) scan spanning from the bottom to the top of the dispersion band, presented for two cases: a single excitation spot on the 15th site -green dots-, and with two excitation spots on the 15th and 17th sites -blue dots-.}
\label{fig:2}
\end{figure} 

Using a spatial light modulator, we focus two coherent laser beams of equal intensity and phase (\( \Delta\phi = 0 \)) on lattice sites \( j-1 \) and \( j+1 \), sketched in Fig.~\ref{fig:2}(a). 
From Eq.~(\ref{eq:analytical}) it can be shown that in the absence of next-nearest-neighbor hopping (\(t'\)), a steady state fully localized at site $j$ can only happen at \( \Delta/t = 0 \) (see Ref.~\cite{Supplementary_arxiv}). 
However, in the experiment, we observe localization at the central site for a laser detuning \( \Delta/t = -0.21 \), corresponding to an energy detuning of \( E_p - E_0 = - 0.07\,\text{meV} \) as displayed in Fig.~\ref{fig:2}(b)-(c). This deviation from $\Delta/t = 0$ is due to the influence of \( t' \), and it is confirmed numerically. At other laser detunings, Fig.~\ref{fig:2}(b)-(c) shows a broader distribution of light across the lattice.
In panel (c) the measured intensity is normalized such that the total sum of the squared intensities over all lattice sites is equal to 1 (i.e., \({\sum_{m} |\psi_{C,m}|^2} = 1\)).

\begin{figure*}[t!]
\centering
\includegraphics[width=\textwidth]{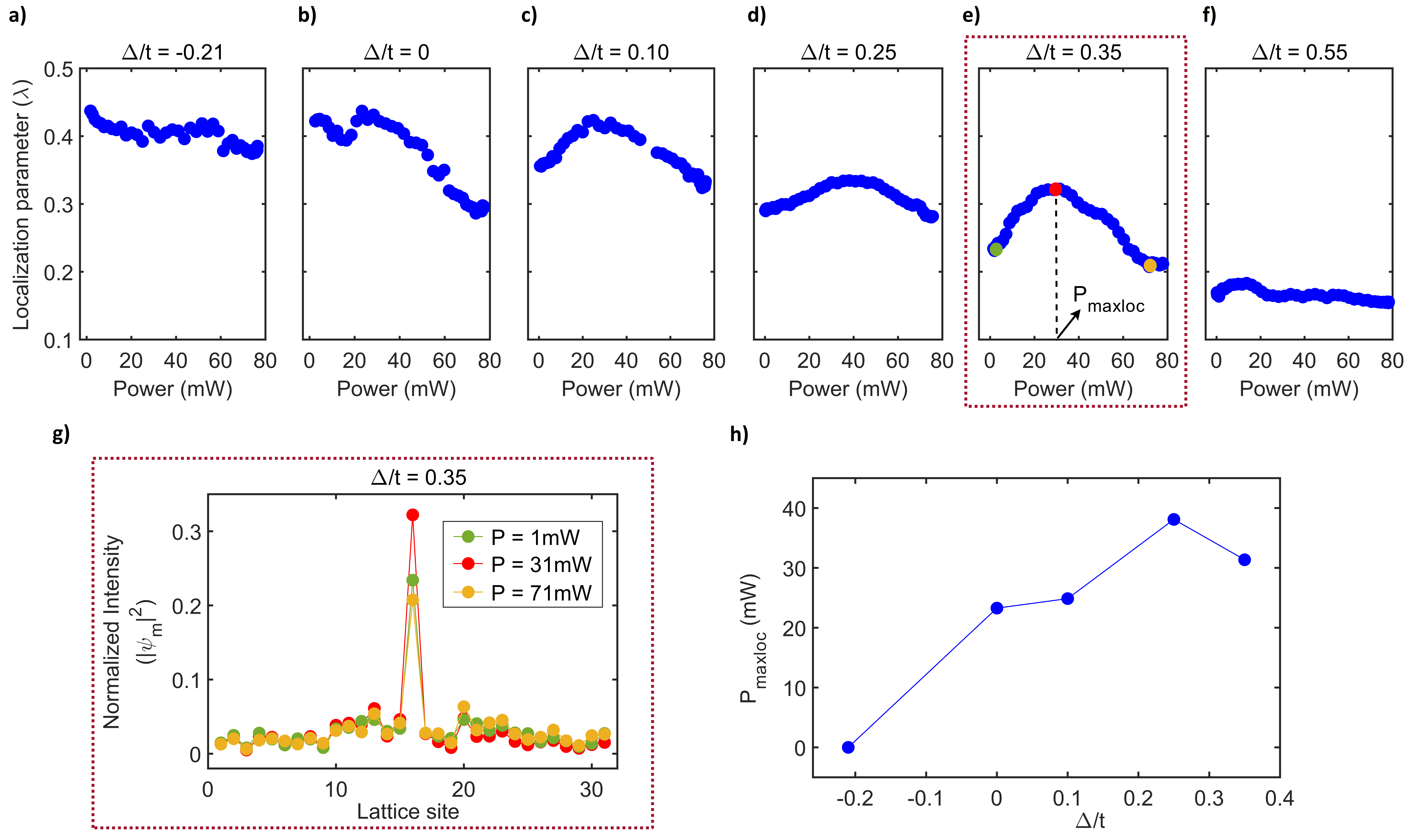} 
\caption{\label{fig:3} (a)-(f) Power dependence of the localization parameter for six different detunings $\Delta/t$: -0.21, 0, 0.10, 0.25, 0.35, and 0.55. 
(g) Line profiles for three different powers (1~mW, 31~mW, and 71~mW) at a detuning of \(\Delta/t = 0.35\), indicated by different markers in (e). (h) Maximum localization power (\( P_{\text{maxloc}} \)) as a function of the first five detunings. For \(\Delta/t = 0.55\), \( P_{\text{maxloc}} \) is not included in (h) as nonlinearity-induced localization is no longer observed.}
\end{figure*}

The degree of localization can be quantified with the localization parameter:
\begin{equation}
\lambda = \dfrac{|\psi_{j}|^2}{\sum_m |\psi_m|^2}.
\label{eq:lambda}
\end{equation} 
It measures the fraction of the total intensity at site $j$ (in between the two pump spots), which is where we expect the localization to happen.
A higher value of \(\lambda\) indicates a greater degree of localization between the pumping spots, whereas a lower value describes extended modes across the lattice.
Figure~\ref{fig:2}(d) -blue dots- displays the measured values of \(\lambda\) in the configuration of panel (a) when the laser energy is scanned across the entire band.
It confirms that maximum localization occurs exclusively at \( \Delta/t = -0.21 \). 
This behavior is very different to the case of a single spot excitation, in which light spreads over all lattice sites at all excitation energies as shown in green dots in Fig.~\ref{fig:2}(d).

\section{Non-Linear localization in extended modes}

In the linear regime, the maximum localization occurs at \( \Delta/t = -0.21 \). 
To understand how the localization is modified by interactions, we explore the role of on-site Kerr nonlinearities in shaping the steady-state field at the site located between the two pump spots. 
A recent theoretical study by Muñoz de las Heras et~al.~\cite{de2024nonlinearity} demonstrated that in the presence of nonlinearities, strong localization can also occur at different detunings. In our experiment, the nonlinear interaction strength is controlled via the power of the coherent pumps, which modulates the polariton density.

Figure~\ref{fig:3}(a)-(f) display the measured localization parameter $\lambda$ as a function of pump power at six laser detunings. 
At \(\Delta/t = -0.21\), at low power, we measure $\lambda= 0.44$, which indicates a strong localization. When increasing the pump power, the localization level slightly decreases.
When the laser detuning is increased to $\Delta/t =0$ (Fig.~\ref{fig:3}(b)), the value of lambda at low power is smaller than at $\Delta/t = -0.21$, and the highest localization is observed at a laser intensity of about 23~mW before significantly declining at higher powers. 
This trend becomes clearer at higher detunings up to $\Delta/t = 0.35$, displayed in Fig~\ref{fig:3}(e), when the highest measured localization takes place at 31mW (see Fig.~\ref{fig:3}(g) for a comparison of the mode distribution at three different powers).
The power at which the highest localization takes place increases with the detuning as displayed in Fig.~\ref{fig:3}(h).
At even higher detunings ($\Delta/t = 0.55$, Fig.~\ref{fig:3}(f)) the measured impact of the nonlinearity becomes weak.

\begin{figure*}[t!]
\centering
\includegraphics[width=\textwidth]{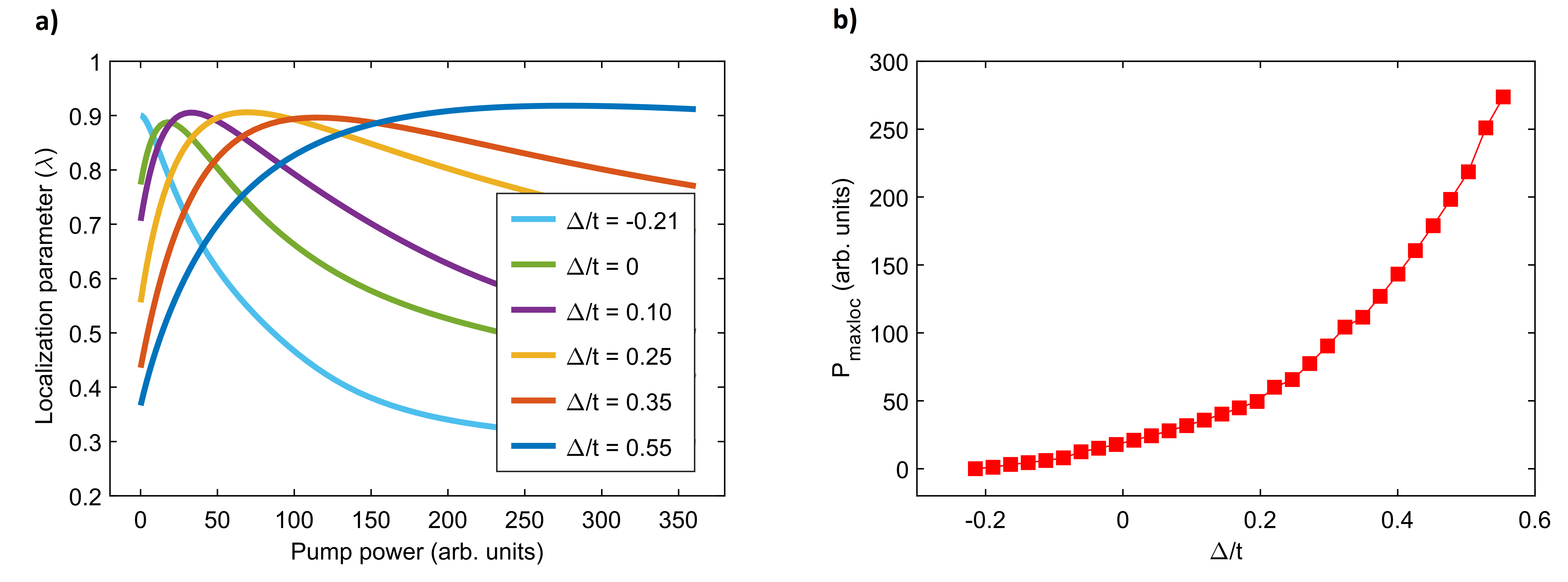} 
\caption{\label{fig:4} (a) Numerical results for the \(\lambda\) dependence on pump power for the same detuning settings as in the experiments in Fig.~\ref{fig:3}, with \(g_x \neq 0 \). 
(b) Maximum localization power (\( P_{\text{maxloc}}\)), as a function of detuning \(\Delta/t\).}
\end{figure*}


The modifications of \(\lambda\) across varying pump powers and laser detunings show that nonlinearity modifies the interference conditions in the lattice to partially re-establish localization.
To further understand this phenomenon we refer to Ref.~\cite{de2024nonlinearity}, which uses a Gross-Pitaevskii model to describe the localization in a lattice similar to the one discussed in our work. 
The model in that work assumes no next-nearest neighbor hopping and it predicts a linear scaling of the power at which maximum localization occurs ($P_{\text{maxloc}}$) with increasing detuning. 
However, this mean-field approach does not account for nonlinear saturation effects that arise from the mixed light-matter nature of polaritons. 
To address this limitation, we use the model presented in Eqs. (\ref{eq:photon}) and (\ref{eq:exciton}). 
Figure~\ref{fig:4}(a) displays the numerically computed value of \(\lambda\) as a function of pump power for various detunings using the lattice parameters of the simulations in Fig.~\ref{fig:1}. 
For \(\Delta/t = -0.21\), the localization is initially high at low pump powers but gradually decreases with increasing power. 
In contrast, for other detunings, \(\lambda\) increases with pump power, reaching a peak at a specific value \(P_{\text{maxloc}}\), after which it begins to decline.
For large detunings such as \(\Delta/t = 0.35\) and \(0.55\), the localization grows more gradually and eventually saturates, reflecting a nonlinear plateau. 
We have verified that including an additional nonlinear term in Eqs.~(\ref{eq:photon})-(\ref{eq:exciton}) to account for the saturation of the oscillator strength~\cite{frerot2023bogoliubov, richard_excitonic_2025} does not qualitatively change the behavior observed in the simulations of Fig.~\ref{fig:4}, even when the oscillator strength saturation dominates over exciton-exciton interaction $g_X$.

These simulations show that a high degree of localization can be recovered at any of the studied detunings for the proper value of driving power. 
However, in contrast to the results in Ref.~\cite{de2024nonlinearity}, the pump power required to reach the maximum $\lambda$ increases superlinearly with detuning, as observed in Fig.~\ref{fig:4}(b). 
The reason is that in the two coupled equation model, exciton interactions lead to an increase of the exciton self-energy, while the photon energy is not affected by them.
As a result, the lower-branch polaritons constituting the lattice modes become increasingly photonic with power, leading to a reduction in their effective interactions.
This self-limiting behavior leads to the superlinear increase of $P_{\text{maxloc}}$ with detuning in  Fig.~\ref{fig:4}(b), and it also smoothens the response with power at high detunings (Fig.~\ref{fig:4}(a)).  
 
While the model displays an increase of $P_{\text{maxloc}}$ with power as in the experiments, there are discrepancies between the simulations in Fig.~\ref{fig:4} and the experiments reported in Fig.~\ref{fig:3}.
In particular, in the experiments, the nonlinear response at large detunings does not allow us to recover the highest localization value observed at $\Delta/t =-0.21$ in the linear regime.
To better understand these differences, we examined several possible contributing factors. 
Previous studies have shown that disorder can disrupt interference-based localization in photonic lattices by degrading coherent transport or suppressing nonlinear localization mechanisms~\cite{lodahl2004controlling,lahini2008anderson, lahini2009observation}, motivating us to assess its impact in our system. 
Therefore, we performed numerical simulations to study the effect of disorder in both the coupling strengths and on-site energies.
Reference~\cite{Supplementary_arxiv} shows a detailed analysis of disorder realizations, including comparisons of the localization parameter and steady-state intensity profiles with and without disorder.
Remarkably, the localization effect remained qualitatively robust with disorder values up to the hopping strength, suggesting that our experiments are stable against such imperfections.
This robustness arises due to the presence of dissipation, which plays a crucial role in protecting localized states from the system’s imperfections. 

\begin{figure}[t]
\centering
\includegraphics[width=\columnwidth]{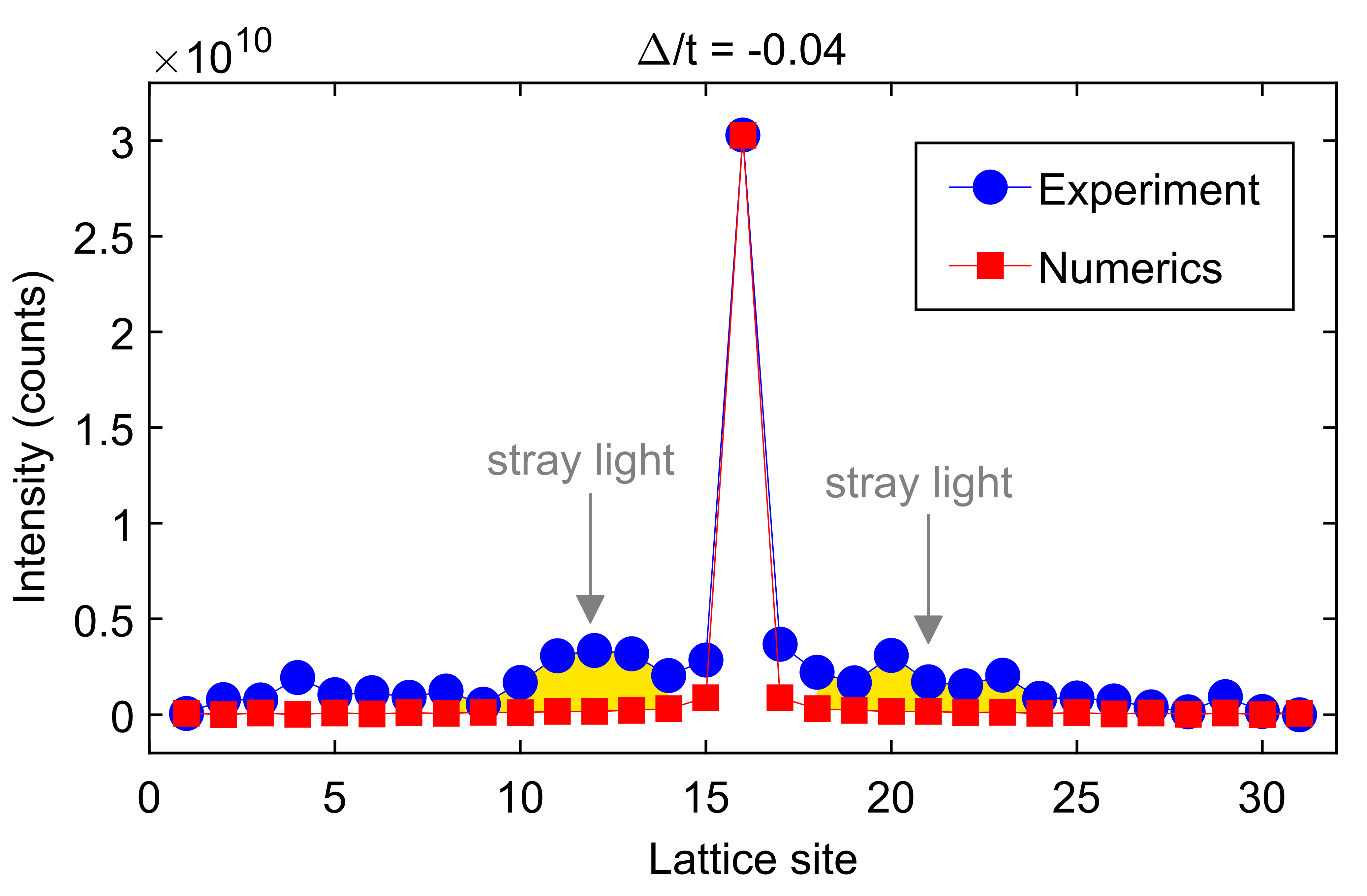}
\caption{Line profile showing the intensity counts for both the experiment and numerical simulations at \(\Delta/t = -0.04\), highlighting the presence of stray light (yellow shadow).}
\label{fig:5}
\end{figure} 

This led us to consider other possible sources of deviation, particularly experimental limitations. 
Stray light is a significant factor that can influence the measured localization parameter: unwanted reflections or scattering from the edges of the microstructures reaching the detection camera may introduce parasitic signals, reducing both the precision and contrast of the measurement. 
Figure~\ref{fig:5} shows the measured intensity profile (blue points) for \(\Delta/t = -0.04\) at low power, along with a simulation in the same conditions (red points). Both sets are normalized to the emitted intensity at the localized site.
A distinct feature in the experimental data is the presence of a bump on both sides of the pumped sites, which is completely absent in the numerical simulations. 
It is highlighted with yellow shadows in Fig.~\ref{fig:5}.
Similar bumps are observed at all powers and detunings in the experiment, which strongly suggests that its origin is stray light.
Indeed, the one-dimensional lattice studied in this work was fabricated via deep etching of the upper and lower mirrors of the microcavity down to the GaAs substrate.
This procedure ensures a deep confinement of the photonic modes inside the micropillar but allows residual laser light to go through the substrate towards the CCD detector and pollute the measured intensity profiles, in particular close to the excitation spots.
Since the localization parameter \( \lambda \) is derived from intensity measurements, the presence of stray light can parasitically contribute to the signal, thereby affecting the extracted value of \(\lambda \). 
This additional contribution may lead to deviations in the observed trend of \( P_{\text{maxloc}}\) between the experimental and numerical results.

\section{Conclusion and Outlook}
To sum up, we have demonstrated that precise control over the laser phase at individual sites and pump laser energy enables the manipulation of interference patterns in driven-dissipative photonic lattices. 
We used this control to experimentally demonstrate the existence of optical switching of ON and OFF responses, as well as directional transport with high efficiency. 
In addition, we investigated the role of interactions, which lead to light confinement in frequency regimes not accessible under linear conditions as proposed in Ref.~\cite{de2024nonlinearity}. 
The observed effects underscore the significant role of interference in presence of nonlinearities and call for further studies in lattices with other band structures~\cite{usaj_localization_2024, real2024controlling} and in the presence of topological properties~\cite{pernet2022gap, gonzalez-tudela_connecting_2022}.

\begin{acknowledgments}
This work was supported by the QUANTERA project MOLAR (PCI2024-153449) and funded by MICIU/AEI/10.13039/501100011033, the Agence National de la Recherche and by the European Union. It has also been supported by the European Union’s Horizon 2020 research and innovation programme through the ERC projects EmergenTopo (grant number no. 865151) and ARQADIA (grant agreement no. 949730); the Marie Skłodowska-Curie grant agreement no.101108433; and under Horizon Europe research and innovation programme  through the ERC project ANAPOLIS (grant agreement no. 101054448).
It was also funded by the French government through the Programme Investissement d'Avenir (I-SITE ULNE /ANR-16-IDEX-0004 ULNE) managed by the Agence Nationale de la Recherche, the Labex CEMPI (ANR-11-LABX-0007), the CDP C2EMPI project (R-CDP-24-004-C2EMPI), as well as the French State under the France-2030 programme, the University of Lille, the Initiative of Excellence of the University of Lille, the European Metropolis of Lille. It was partly supported by the Paris Ile de France R\'egion in the framework of DIM SIRTEQ and DIM QuanTIP and by the RENATECH network and the General Council of Essonne.
BR acknowledges ANID Fondecyt de postdoctorado No 3230139.
AGT acknowledges support from the CSIC Research Platform on Quantum Technologies PTI-001 and from Spanish projects PID2021-127968NB-I00 funded by MICIU/AEI/10.13039/501100011033/ and by FEDER Una manera de hacer Europa, TED2021-130552B-C22 funded by  MICIU/AEI /10.13039/501100011033 and by the European Union NextGenerationEU/ PRTR.
AMH acknowledges support from Fundación General CSIC’s ComFuturo program, which has received funding from the European Union’s Horizon 2020 research and innovation program under the Marie Skłodowska-Curie Grant Agreement No. 101034263.

\end{acknowledgments}



%

\end{document}